\documentclass[a4paper,11pt]{article}
\usepackage{jheppub} 
\usepackage[T1]{fontenc} 
\usepackage{multirow}
\usepackage{color}
\usepackage{ulem}
\usepackage{rotating}
\usepackage{booktabs}
\usepackage{dcolumn}
\usepackage{array} 
\usepackage{multirow}
\usepackage{lineno} 
\usepackage{siunitx}
\usepackage{amsmath}
\usepackage{bm}
\usepackage{mathrsfs}
\usepackage{adjustbox}

\DeclareSIUnit{\bmm}{\bm{m}}

\DeclareSIUnit{\clight}{\textnormal{\textit{c}}}

\newcolumntype{d}{D{.}{.}{-1}}
\newcolumntype{e}{D{.}{.}{8}}
\newcolumntype{f}{D{.}{.}{18}}
\newcolumntype{h}{D{.}{.}{13}}
\newcolumntype{g}{D{.}{.}{12}}

\newcommand{\BB}{B\bar{B}}
\newcommand{\XXB}{\Xi^{-}\bar\Xi^{+}}
\newcommand{\EE}{e^+e^-}
\newcommand{\ar}{\rightarrow}
\newcommand{\bbt}{\bibitem}

\title{\protect\boldmath Measurement of the cross section of $\EE\ar\Xi^{-}\bar\Xi^{+}$ at center-of-mass energies between $\mathbf{3.510}$ and \si{\mathbf{4.843}\,{\textbf{GeV}}}}

\collaborationImg{\includegraphics[width=0.15\textwidth]{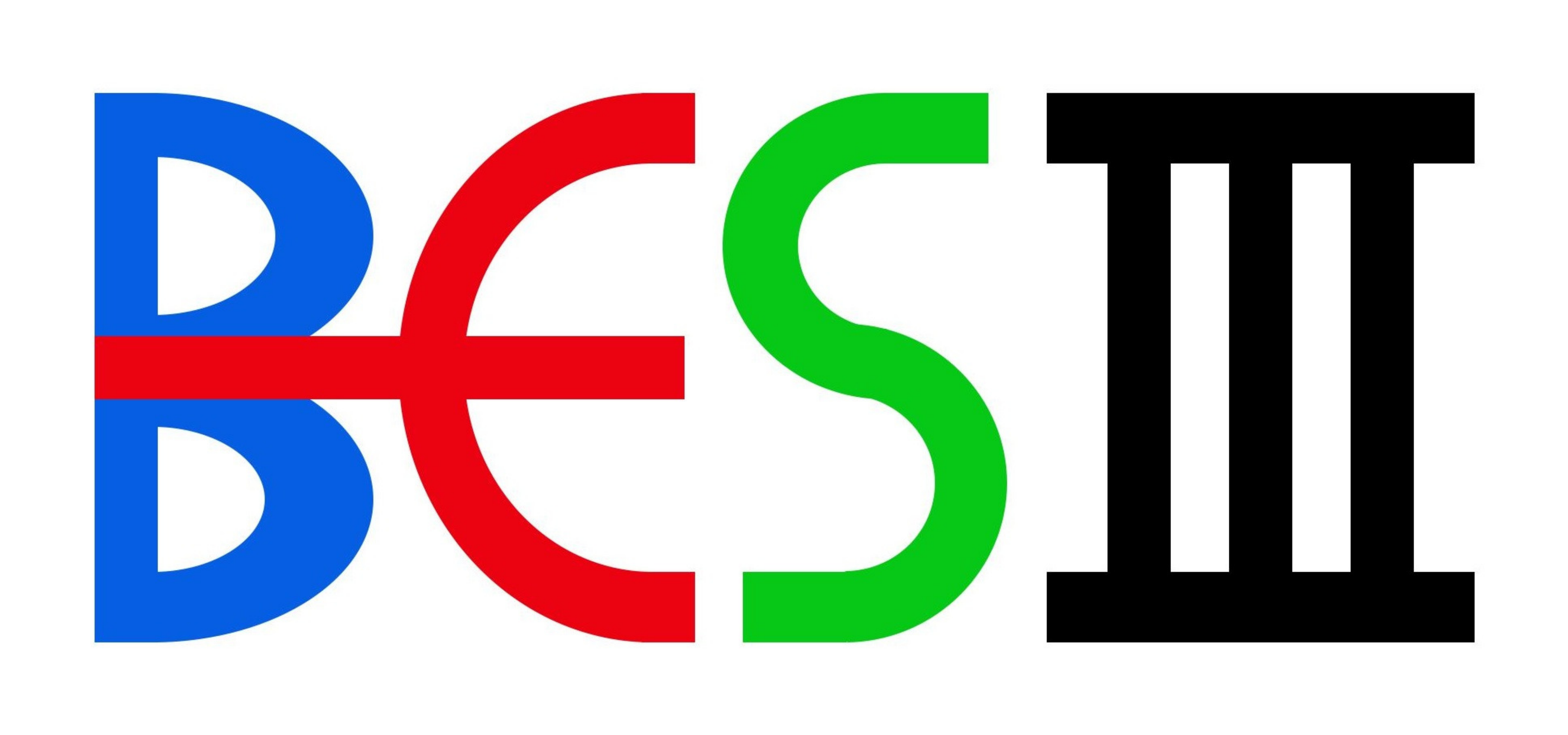}}
\collaboration{The BESIII Collaboration}

\author{
M.~Ablikim$^{1}$, M.~N.~Achasov$^{5,b}$, P.~Adlarson$^{74}$, X.~C.~Ai$^{80}$, R.~Aliberti$^{35}$, A.~Amoroso$^{73A,73C}$, M.~R.~An$^{39}$, Q.~An$^{70,57}$, Y.~Bai$^{56}$, O.~Bakina$^{36}$
% ...
}
\affiliation{
$^{1}$ Institute of High Energy Physics, Beijing 100049, People's Republic of China\\
% ...
$^{5}$ Budker Institute of Nuclear Physics SB RAS (BINP), Novosibirsk 630090, Russia\\
% ...
$^{35}$ Johannes Gutenberg University of Mainz, Johann-Joachim-Becher-Weg 45, D-55099 Mainz, Germany\\
$^{36}$ Joint Institute for Nuclear Research, 141980 Dubna, Moscow region, Russia\\
% ...
$^{39}$ Liaoning Normal University, Dalian 116029, People's Republic of China\\
% ...
$^{56}$ Southeast University, Nanjing 211100, People's Republic of China\\
$^{57}$ State Key Laboratory of Particle Detection and Electronics, Beijing 100049, Hefei 230026, People's Republic of China\\
% ...
$^{70}$ University of Science and Technology of China, Hefei 230026, People's Republic of China\\
% ...
$^{73}$ University of Turin and INFN, (A)University of Turin, I-10125, Turin, Italy; (B)University of Eastern Piedmont, I-15121, Alessandria, Italy; (C)INFN, I-10125, Turin, Italy\\
$^{74}$ Uppsala University, Box 516, SE-75120 Uppsala, Sweden\\
% ...
$^{80}$ Zhengzhou University, Zhengzhou 450001, People's Republic of China\\

\vspace{0.2cm}
% ...
$^{b}$ Also at the Novosibirsk State University, Novosibirsk, 630090, Russia\\
% ...
}
\emailAdd{besiii-publications@ihep.ac.cn}

\begin{document} 

\abstract{
Using $\EE$ collision data corresponding to a total integrated luminosity of \SI{12.9}{fb^{-1}}
collected with the BESIII detector at the BEPCII collider,
the exclusive Born cross sections and the effective form factors of
the reaction $\EE\ar\XXB$ are measured via the single baryon-tag method
at 23 center-of-mass energies between 3.510 and \SI{4.843}{GeV}.
Evidence for the decay $\psi(3770)\ar\XXB$ is observed with a
significance of 4.5$\sigma$ by analyzing the measured cross sections
together with earlier BESIII results. For the other charmonium(-like)
states $\psi(4040)$, $\psi(4160)$, $Y(4230)$, $Y(4360)$, 
$\psi(4415)$, and $Y(4660)$, no significant signal of their decay to $\Xi^-\bar \Xi^+$ is  found. 
For these states, upper limits of the products of the branching fraction and the electronic partial width at the 90\% confidence level are provided. }

\maketitle
\flushbottom
\section{Introduction}
\label{sec:intro}

Below the open-charm threshold, the mass spectrum of the conventional charmonium states is well matched to the predictions from the potential quark model~\cite{Barnes:2005pb}. Above the open-charm threshold, this model predicts five vector charmonium states between the threshold and \SI{4.9}{GeV/{\clight}^{2}}, namely, the $3S$, $2D$, $4S$, $3D$, and $5S$ states. 
However, an overpopulation of vector states has been observed in this energy region. 
Three of them, $\psi(4040)$, $\psi(4160)$, $\psi(4415)$, are dominated by open-charm final states \cite{BES:2001ckj}.
Other states, e.g. $Y(4230)$, $Y(4260)$, $Y(4360)$, $Y(4634)$, $Y(4660)$, are mainly observed in hidden-charm final states, 
produced via initial state radiation (ISR) processes at \textsc{BaBar} and 
Belle~\cite{BaBar:2005hhc, Belle:2007dxy, BaBar:2006ait, Belle:2007umv, Belle:2008xmh, BaBar:2012hpr, Belle:2014wyt, BaBar:2012vyb, Belle:2013yex} 
or via direct production processes at  CLEO~\cite{CLEO} and BESIII~\cite{BESIIIAB, BESIII:2023cmv}.
These $Y$ states cannot be easily accommodated in the charmonium spectrum predicted in the quark model, and many hypotheses have been proposed to interpret them~\cite{Farrar,Briceno,Chen:2016qju,Close:2005iz, Qian:2021neg, Bai:2023dhc, Xia:2015mga,Yan:2023yff,Dai:2023vsw}, such as hybrids, multiple-quark states, molecules,  \textit{etc.} However, no solid conclusion can be made at present and puzzles remain for these vector charmonium(-like) states. 

This situation reflects the present insufficient understanding of the strong interaction, especially in the non-perturbative regime. 
To improve the situation, additional experimental measurements are desirable. Among these measurements, the decays of $\psi/Y \to$ 
light baryon-antibaryon$(B \bar{B})$ are promising due to the simple topologies of the final states and relatively well understood mechanisms, dominated by three-gluons or one-photon processes.
Moreover, measurements of electromagnetic form factors and effective form factors 
of the $B\bar{B}$ pairs may provide insight into the internal structure of the charmonium(-like) states. Up to now the
experimental information on $B\bar{B}$ decays of these charmonium-(like) states is scarce above the open-charm threshold.
Although many experimental 
studies~\cite{BESIII:2021ccp, Ablikim:2013pgf, Belle:2008xmh, Ablikim:2019kkp, Liu:2023xhg,
BESIII:2017kqg, Wang:2021lfq, Wang:2022bzl} 
of $B\bar{B}$ pair production are performed in this energy region by the BESIII and Belle experiments, 
no $B\bar{B}$ decays of the vector charmonium(-like) states 
have been observed except for the evidence of $\psi(3770)\to\Lambda\bar\Lambda$.
It could indicate that the production of $\XXB$ mainly proceeds through the continuum process and not through charmonium resonances, justifying the assumption of one-photon production.
Thus, more precise measurements of exclusive cross sections of $e^+e^-\to\BB$   final states above the open-charm threshold could provide 
a valuable information to understand the nature of these vector charmonium(-like) states.

In this article, measurements of the Born cross section and the effective form factor for the process $\EE\ar\XXB$ are presented using data corresponding to a total integrated luminosity of \SI{12.9}{fb^{-1}} collected at center-of-mass (CM) energies $\sqrt{s}$ between 3.510 and \SI{4.843}{GeV} with the BESIII detector~\cite{besiii} at the BEPCII collider~\cite{BEPCII}. In addition, potential contributions from resonances are studied by fitting the dressed cross section of the $\EE\to\XXB$ reaction combined with the earlier BESIII results reported in 2020~\cite{Ablikim:2019kkp}.

\section{BESIII Detector and Monte Carlo simulation}
The BESIII detector~\cite{besiii} records symmetric $e^+e^-$ collisions 
provided by the BEPCII storage ring~\cite{BEPCII}
in the CM energy range from 2.0 to \SI{4.95}{GeV},
with a peak luminosity of \SI{1e33}{\per\centi\meter\squared\per\second}
achieved at $\sqrt{s} =$ \SI{3.77}{GeV}. 
BESIII has collected large data samples in this energy region~\cite{Ablikim:2019hff, EcmsMea, EventFilter}. The cylindrical core of the BESIII detector covers 93\% of the full solid angle and consists of a helium-based
 multilayer drift chamber~(MDC), a plastic scintillator time-of-flight
system~(TOF), and a CsI(Tl) electromagnetic calorimeter~(EMC),
which are all enclosed in a superconducting solenoidal magnet
providing a \SI{1.0}{T} magnetic field. The solenoid is supported by an
octagonal flux-return yoke with resistive plate counter muon
identification modules interleaved with steel. 
The charged-particle momentum resolution at \SI{1}{GeV/\clight} is
$0.5\%$, and the 
${\rm d}E/{\rm d}x$
resolution is $6\%$ for electrons
from Bhabha scattering. The EMC measures photon energies with a
resolution of $2.5\%$ ($5\%$) at \SI{1}{GeV} in the barrel (end cap)
region. The time resolution in the TOF barrel region is \SI{68}{ps}, while
that in the end cap region is \SI{110}{ps}. The end cap TOF
system was upgraded in 2015 using multigap resistive plate chamber
technology, providing a time resolution of \SI{60}{ps}, which benefits $77\%$ of the data used in this analysis~\cite{etof1, etof2, etof3}.

Simulated data samples produced with a {\sc geant4}-based~\cite{GEANT4} Monte Carlo (MC) package, 
which includes the geometric description of the BESIII detector~\cite{Huang:2022wuo} and the detector response, 
are used to determine the signal detection efficiency. 
A sample of 200,000 $\EE\to\XXB$ events 
is simulated with a uniform phase space (PHSP) distribution for each of the 23 CM energy points in the range from 3.510 to \SI{4.843}{GeV}. 
The decay chain of the $\Xi^{-}$ baryon is handled by the {\sc evtgen} program~\cite{evtgen2,EVTGEN} using a PHSP model. 
The production process is simulated by the {\sc kkmc} generator \cite{KKMC} that includes the beam energy spread 
and ISR in the $e^+e^-$ annihilation.

\section{Event selection}
The selection of $e^{+}e^{-} \to \XXB$ events with a full reconstruction method suffers from low reconstruction efficiency. Therefore, a partial reconstruction technique is employed, in which only the $\Xi^-$ baryon is reconstructed via its decay $\Xi^-\to\Lambda\pi^-$ with the subsequent decay $\Lambda\to p\pi^-$, and the presence of the $\bar{\Xi}^+$ anti-baryon is inferred from the system recoiling against the reconstructed $\Xi^-$. Unless otherwise noted, the charge conjugated state is implicitly included throughout this paper.

Tracks of charged particles detected in the MDC are required to be within the optimal MDC angular coverage, $|\cos\theta| < 0.93$, where the polar-angle $\theta$ is defined with respect to the MDC symmetry axis. At least one positively charged and two negatively charged tracks are required, which are required to be well reconstructed in the MDC with good helix fits. 
The particle identification (PID) for charged particles combines measurements of the d$E$/d$x$ by the MDC and the time of flight by the TOF to form likelihoods ${\cal L}(h)$ $(h = p, K, \pi)$ for each hadron hypothesis $h$. 
Tracks are identified as protons if the proton hypothesis has the greatest likelihood $({\cal L}(p) > {\cal L}(K)$ 
and ${\cal L}(p) > {\cal L}(\pi))$, while charged pions are identified by requiring that
${\cal L}(\pi) > {\cal L}(p)$ and ${\cal L}(\pi) > {\cal L}(K)$. Only events with at least two negatively charged pions and one proton are kept for the further analysis.

The reconstruction of $\Lambda$ and $\Xi^-$ baryons follows Refs.
~\cite{BESIII:2012ghz,  BESIII:2021cvv,  BESIII:2023euh,BESIII:2022kzc}. Briefly, to reconstruct $\Lambda$ candidates and suppress non-$\Lambda$ background, a secondary vertex fit~\cite{vtxfit} is applied to all $p\pi^-$ combinations, and their invariant mass is required to be within \SI{5}{MeV/\clight^{2}} from the known $\Lambda$ mass.  The selection window is determined by optimizing a figure of merit (FOM) defined as $S/\sqrt{S+B}$, where $S$ is the number of signal MC events and $B$ is the number of the background events expected from simulation. 
Similarly, the $\Xi^{-}$ candidates are reconstructed using a further secondary vertex fit applied
to the combination of the selected $\pi^-$ and $\Lambda$ candidates which minimizes the difference $|M_{\pi^-\Lambda}-m_{\Xi^-}|$, where $M_{\pi^-\Lambda}$ is the invariant mass of the $\pi^-\Lambda$ combination and $m_{\Xi^-}$ is the known mass of $\Xi^{-}$ from the Particle Data Group (PDG)~\cite{PDG2020}. 
The signal window is defined by requiring that $M_{\pi^-\Lambda}$ is within \SI{10}{MeV/\clight^{2}} from the known $\Xi^{-}$ mass.  This width of the window is determined by optimizing the FOM and is about three times the experimental mass resolution. 
The decay length of the $\Xi^{-}$ candidate, \textit{i.e.}
the distance between the average position of the $e^+e^-$ collisions and the decay vertex of $\Xi^{-}$, 
is required to be greater than zero.
The decay length of the $\Lambda$ candidate is also required to be greater than zero.

\begin{figure}[!hbpt]
\begin{center}
\includegraphics[width=1.0\textwidth]{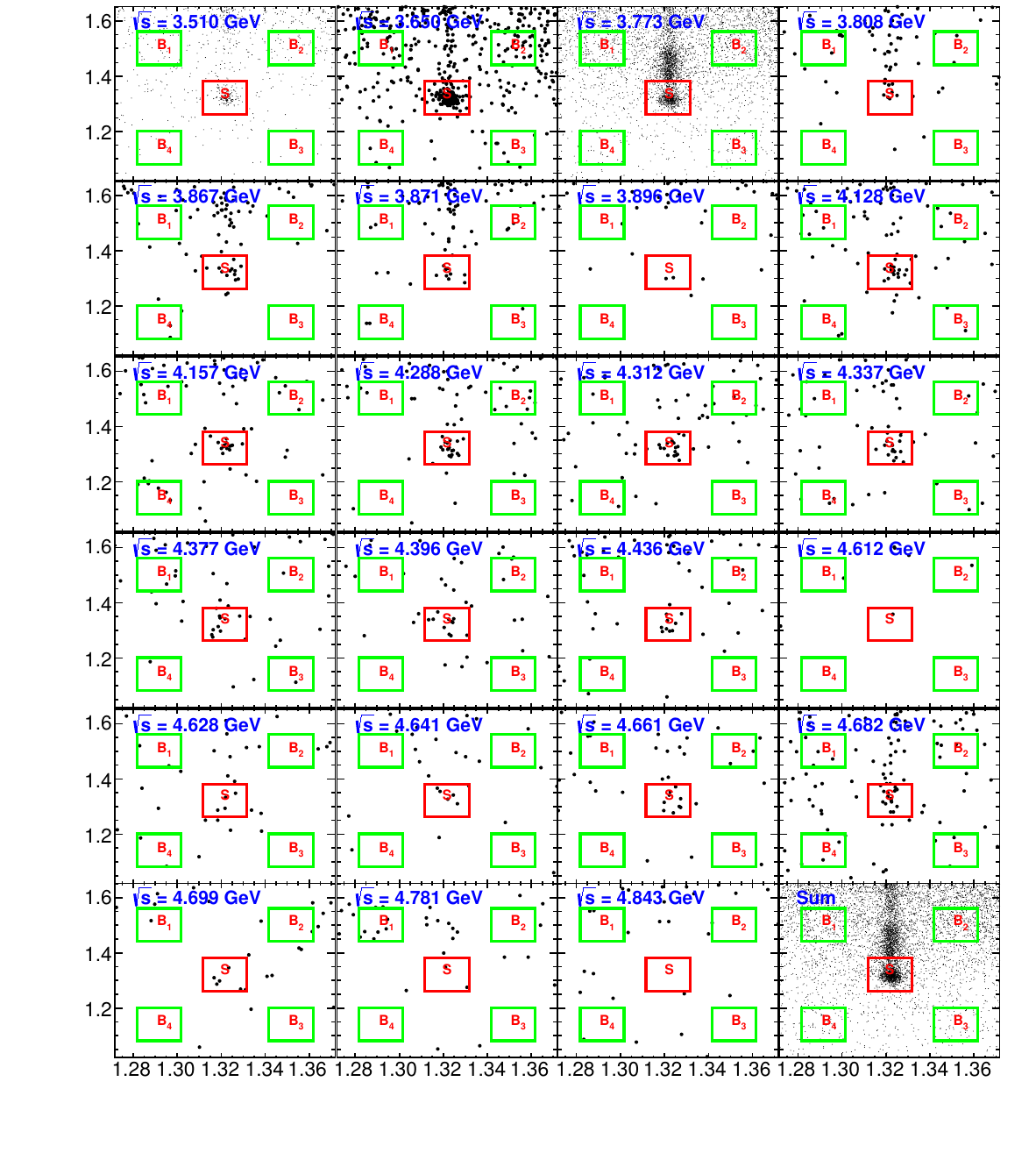}
\put(-230, 10){\boldmath $M_{\pi^{-}\Lambda}$ \textbf{(GeV)}}
\put(-425, 220){\rotatebox{90}{\boldmath $M^{\rm recoil}_{\pi^{-}\Lambda}$ \textbf{(GeV)}}}
\end{center}
\caption{
Distributions of $M_{\pi^{-}\Lambda}$ versus $M^{\rm recoil}_{\pi^{-}\Lambda}$ for each energy point and for the sum of all energy points (bottom, right) between 3.510 and \SI{4.843}{GeV} from data.
The red boxes represent the signal regions and the green boxes represent the selected sideband regions.}
\label{Fig:SUM:DATA}
\end{figure}

To extract anti-baryon $\bar\Xi^+$ candidates, we calculate the mass recoiling against the selected $\pi^-\Lambda$ system as
\begin{equation}
	M^{\rm recoil}_{\pi^-\Lambda} = \sqrt{(\sqrt{s}-E_{\pi^-\Lambda})^{2} - \vec{p}^{2}_{\pi^-\Lambda}},
\end{equation}
where $E_{\pi^-\Lambda}$ and $\vec{p}_{\pi^-\Lambda}$ are the energy and momentum of the selected $\pi^-\Lambda$ candidates in the CM frame, and $\sqrt{s}$ is the CM energy. 
The signal window is defined by requiring that $M_{\pi^-\Lambda}^{\rm recoil}$ is within \SI{60}{MeV/\clight^{2}} from the known $\Xi^{-}$ mass.
Figure~\ref{Fig:SUM:DATA} shows the scatter plots of  $M_{\pi^-\Lambda}$ versus $M^{\rm recoil}_{\pi^-\Lambda}$ for each energy point separately and for the sum of all energy points. A clear cluster of events around the nominal $\Xi^{-}$ mass can be seen.

\section{Born cross section measurement}
\subsection{Extraction of signal yields}
After applying the above selection criteria, the surviving  background events are mainly from non-$\XXB$ events, such as $\EE\to\pi^{+}\pi^{-}\Lambda\bar\Lambda$.
The background yield in the signal region is estimated using four sideband regions
$B_{i}$, where $i = 1$, 2, 3, 4, each with the same area as the signal region. The regions are shown in Figure~\ref{Fig:SUM:DATA}, and the exact ranges are defined by
\begin{itemize}
 \item $B_{1}$: $M_{\pi^-\Lambda}\in$ [1.282, 1.302] GeV/$c^2$ $\&$ $M^{\rm recoil}_{\pi^-\Lambda}\in$ [1.442, 1.562] GeV/$c^2$, 
 \item $B_{2}$: $M_{\pi^-\Lambda}\in$ [1.342, 1.362] GeV/$c^2$ $\&$ $M^{\rm recoil}_{\pi^-\Lambda}\in$ [1.442, 1.562] GeV/$c^2$, 
 \item $B_{3}$: $M_{\pi^-\Lambda}\in$ [1.342, 1.362] GeV/$c^2$ $\&$ $M^{\rm recoil}_{\pi^-\Lambda}\in$ [1.082, 1.202] GeV/$c^2$, 
 \item $B_{4}$: $M_{\pi^-\Lambda}\in$ [1.282, 1.302] GeV/$c^2$ $\&$ $M^{\rm recoil}_{\pi^-\Lambda}\in$ [1.082, 1.202] GeV/$c^2$.
\end{itemize}
The signal yield $N_{\rm obs}$ for the $e^{+}e^{-}\rightarrow\XXB$ reaction at each energy point can then be determined by subtracting the number of events in the sideband regions from the signal region, i.e. $N_{\rm obs} = N_{S} - N_{\rm bkg}$, where $N_{S}$ is number of events in the signal region, $N_{\rm bkg}=\frac{1}{4}\sum^{4}_{i=1}B_{i}$.  The signal yields are listed in Table~\ref{tab:signal:yields:DD01}.   
The single baryon-tag method leads to a double counting effect for the $\XXB$ final state, which is taken into account in the calculation of the statistical uncertainty based on the study of MC simulation~\cite{Ablikim:2019kkp}. This does not affect the central value of the final result but it requires a correction of the statistical uncertainty, which is underestimated by approximately $12\%$.

\begin{table}[!hpt]
	\begin{center}
	\caption{
Number of signal ($N_{S}$) and background ($N_{\rm bkg}$) events for each energy points.  The upper limits ($N^{\rm UL}$) at the 90\% confidence level including the systematic uncertainty, and the statistical uncertainties for the observed events are determined by the TRolke method with Poisson background and Gaussian efficiency  model~\cite{Lundberg:2009iu}. The signal significance ($\mathcal{S}$) is estimated as $N_{S}/\sqrt{N_{S}+N_{\rm bkg}}$.} 
	{
	\begin{tabular}{cdegd} \hline \hline
\multicolumn{1}{c}{$\sqrt{s}$ (GeV)} & \multicolumn{1}{c}{$N_{S}$\,\,\,} & \multicolumn{1}{l}{\,\,$N_{\rm bkg}$} & \multicolumn{1}{l}{$N_{\rm obs}$ ($N^{\rm UL}$)} & \multicolumn{1}{c}{$\mathcal{S}$ $(\sigma)$} 
\\ \hline

3.510 &143.0 &18.5 &124.5^{+12.8}_{-11.1} &10.4\\
3.650 &144.0 &11.5 &132.5^{+12.8}_{-11.2} &11.0\\
3.773 &703.0 &100.8 &602.2^{+27.7}_{-25.4} &22.7\\
3.808 &8.0 &1.8 &6.2^{+4.0}_{-1.7}~($<$13.2) &2.2\\
3.867 &12.0 &1.0 &11.0^{+3.8}_{-3.1} &3.2\\
3.871 &9.0 &2.0 &7.0^{+3.3}_{-2.7}~($<$13.4) &2.3\\
3.896 &2.0 &0.2 &1.8^{+2.0}_{-0.9}~($<$5.4) &1.2\\
4.128 &18.0 &2.8 &15.2^{+5.4}_{-3.1} &3.6\\
4.157 &14.0 &3.2 &10.8^{+4.3}_{-3.2}~($<$18.8) &2.9\\
4.288 &19.0 &2.2 &16.8^{+4.9}_{-3.8} &3.8\\
4.312 &17.0 &1.5 &15.5^{+5.0}_{-3.3} &3.8\\
4.337 &16.0 &1.8 &14.2^{+5.1}_{-2.9} &3.6\\
4.377 &11.0 &1.5 &9.5^{+4.2}_{-2.5}~($<$16.8) &2.9\\
4.396 &11.0 &0.5 &10.5^{+4.2}_{-2.5} &3.2\\
4.436 &9.0 &1.0 &8.0^{+3.3}_{-2.7}~($<$14.3) &2.7\\
4.612 &1.0 &0.5 &0.5^{+1.9}_{-0.2}~($<$3.7) &0.5\\
4.628 &4.0 &1.2 &2.8^{+2.5}_{-1.5}~($<$7.5) &1.4\\
4.641 &6.0 &0.0 &6.0^{+2.8}_{-2.1}~($<$11.1) &2.4\\
4.661 &9.0 &1.0 &8.0^{+3.3}_{-2.7}~($<$14.3) &2.7\\
4.682 &16.0 &3.8 &12.2^{+5.1}_{-2.9} &3.1\\
4.699 &6.0 &0.5 &5.5^{+3.3}_{-1.6}~($<$11.1) &2.2\\
4.781 &2.0 &1.5 &0.5^{+2.3}_{-0.5}~($<$4.6) &0.4\\
4.843 &0.0 &0.8 &-0.8^{+1.7}_{-0.8}~($<$2.0) &0.0\\

	\hline \hline
	\end{tabular}
	}
	\label{tab:signal:yields:DD01}
	\end{center}
\end{table}

\subsection{Determination of the Born cross section}
The Born cross section for the $\EE\to\XXB$ process at a given CM energy is calculated as
\begin{equation}
\sigma^{B} =\frac{N_{\rm obs}}{2 \cdot {\cal{L}}\cdot(1 + \delta)\cdot\frac{1}{|1 - \Pi|^{2}}\cdot\epsilon\cdot {\cal B}(\Xi^-\ar\pi^-\Lambda)\cdot {\cal B}(\Lambda\ar p\pi^-)},
\end{equation}
where $N_{\rm obs}$ is the number of the observed signal events, a factor of 2 represents the charge conjugate mode included,
${\cal{L}}$ is the integrated luminosity, $(1 + \delta)$ is the ISR correction factor, $\frac{1}{|1-\Pi|^2}$ is the vacuum polarization (VP) correction factor, $\epsilon$ is the detection efficiency, while ${\cal B}(\Xi^-\ar\pi^-\Lambda)$ and ${\cal B}(\Lambda\ar p\pi^-)$ are the corresponding branching fractions taken from the world average~\cite{PDG2020}.
The ISR correction factor is obtained using the QED calculation as described in Ref.~\cite{Kuraev:1985hb}.
The VP correction factor is calculated according to Ref.~\cite{Jegerlehner:2011ti}.
The Born cross section is not well determined beforehand, thus an iterative weighting method for updating the efficiency and ISR correction factor is used as proposed in Ref.~\cite{Sun:2020ehv}. This procedure is repeated until the difference of $(1 + \delta)\epsilon$ between iterations is less than $0.5\%$. The efficiency, ISR correction factor, and Born cross section of the last round are accepted as the final results. 
The measured cross sections for each energy point together with the CLEO-c results at $\sqrt s= 3.770$ and \SI{4.160}{GeV}~\cite{Dobbs:2017hyd} are shown in Figure~\ref{Fig:XiXi::CS:BCS_VS_EFF}.

\section{Determination of the effective form factor}
\begin{figure}[!hbpt]
	\begin{center}
	\includegraphics[width=0.8\textwidth]{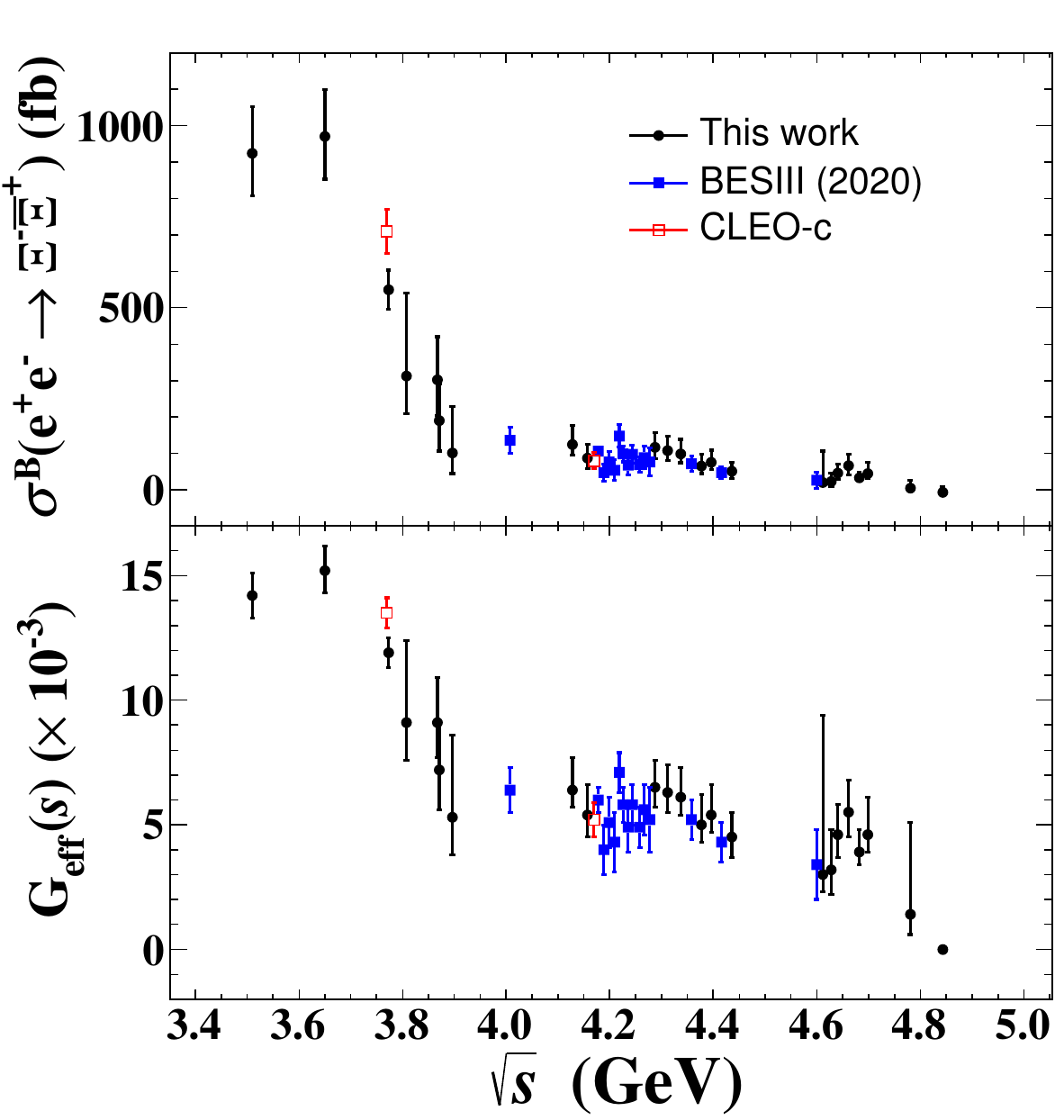}
	\end{center}
	\caption{Measured Born cross section (top) and $\Xi^{-}$ effective form factor (bottom) for $\EE\to\XXB$ as a function of the CM energy, where the uncertainties include both the statistical and systematic ones. }
	\label{Fig:XiXi::CS:BCS_VS_EFF}
\end{figure}

Under the assumption that the dominant process is one-photon exchange $\EE\ar\gamma^{*}\ar\XXB$, one can parameterize the differential cross section in terms of electric and magnetic form factors $G_{E}$ and $G_{M}$.
These are assumed to be continuous functions of the momentum transfer squared, $s=q^{2}$.
The differential cross section can be written~\cite{Baldini} as
\begin{equation}\label{FF01}
\frac{d\sigma^{B}(s)}{d\Omega} = \frac{\alpha^{2}\beta C}{4s} \left[|G_{M}(s)|^{2}(1+\cos^{2}\theta) + \frac{1}{\tau}|G_{E}(s)|^{2}\sin^{2}\theta\right],
\end{equation}
where $s$ is the square of the CM energy, $\alpha$ is the fine structure constant, 
the variable $\beta =\sqrt{1-\frac{1}{\tau}}$ is the velocity, $\tau = \frac{s}{4m^{2}_{\Xi^-}}$,
and the Coulomb correction factor $C$~\cite{Baldini, Arbuzov} parameterizes the electromagnetic interaction between the outgoing baryon and anti-baryon. For neutral baryons, the Coulomb factor is unity, while for point-like charged fermions $C = \frac{\pi\alpha}{\beta}\cdot\frac{\sqrt{1-\beta^2}}{1-e^{-\frac{\pi\alpha}{\beta}}}$~\cite{Sommerfeld, Sakharov,Tzara, Wang:2022zyc}.
Similarly, the Born cross section can be derived by integrating over the full solid angle as
\begin{equation}\label{FF02}
  \sigma^{B}(s) = \frac{4\pi\alpha^{2}C\beta}{3s} \left[|G_{M}(s)|^{2} + \frac{1}{2\tau}|G_{E}(s)|^{2}\right].
\end{equation}
Furthermore, we define the effective form factor $G_{\rm eff}(s)$ as a linear combination of the electric and magnetic form factors as 
\begin{equation}\label{FF04}
|G_{\rm eff}(s)|= \sqrt{
\frac{2\tau|G_{M}(s)|^{2} + |G_{E}(s)|^{2}}{2\tau +1}},
\end{equation}
which can be transformed to
\begin{equation}
|G_{\rm eff}(s)|  = \sqrt{\frac{3s\sigma^{B}}{4\pi\alpha^2C\beta(1+\frac{2m^{2}_{\Xi^-}}{s})}},
\end{equation}
and its uncertainty is propagated to be 
\begin{equation}\label{FF03_err}
\delta_{|G_{\rm eff}(s)|} =\frac{1}{2}C^{\prime}\sqrt{\frac{1}{\sigma^{B}}}\cdot\delta_{\sigma^{B}},
\end{equation}
where $\delta_{\sigma^{B}}$ is the uncertainty of the Born cross section and 

\begin{equation}\label{FF03}
C^{\prime} = \sqrt{\frac{3s}{4\pi\alpha^2C\beta(1+\frac{2m^{2}_{\Xi^-}}{s})}},
\end{equation}
where $m_{\Xi^-}$ is the $\Xi^-$ baryon mass.
The results for the effective form factors at each energy point are shown in Figure~\ref{Fig:XiXi::CS:BCS_VS_EFF} and summarized in Table~\ref{tab:signal:yields:DD}.

\begin{table}[!hpt]
	\begin{center}
	\caption{\small
The CM energy ($\sqrt{s}$), the integrated luminosity (${\cal{L}}$),
the VP correction factor ($\frac{1}{|1 - \Pi|^{2}}$), the ISR correction factor and the detection efficiency ($D=(1 + \delta)\epsilon$), the signal yields ($N_{\rm obs}$), the upper limit of signal yield at the 90\% confidence level ($N^{\rm UL}$), the Born cross section ($\sigma^{B}$), the effective form factor ($|G_{\rm eff}(s)|$) and the statistical significance ($\mathcal{S}$). The first and second uncertainties for $\sigma^{B}$ are statistical and systematic, respectively.}
\scalebox{0.72}{
	
        \begin{tabular}{ldcegfhd} \hline \hline

\multicolumn{1}{c}{$\sqrt{s}$ (GeV)} &\multicolumn{1}{c}{${\cal{L}}$ (pb$^{-1})$} &\multicolumn{1}{c}{\,\,$\frac{1}{|1 - \Pi|^{2}}$} &\multicolumn{1}{l}{$D$ $(\times 10^{-2})$} &\multicolumn{1}{l}{$N_{\rm obs}$ ($N^{\rm UL}$)} &\multicolumn{1}{l}{\;\;\;$\sigma^{B}$ (fb)} &\multicolumn{1}{c}{$|G_{\rm eff}(s)|$ $(\times 10^{-3})$} &\multicolumn{1}{c}{$\mathcal{S}$ $(\sigma)$}

\\ \hline

3.510 &405.7 &1.04 &24.9 \pm 0.1 &124.5^{+12.8}_{-11.1} &923.6^{+106.4}_{-92.3} \pm 70.2 &14.2^{+0.8}_{-0.7} \pm 0.5&10.4\\
3.650 &410.0 &1.02 &25.6 \pm 0.1 &132.5^{+12.8}_{-11.2} &970.8^{+105.1}_{-92.0} \pm 72.8 &15.2^{+0.8}_{-0.7} \pm 0.6&11.0\\
3.773 &2931.8 &1.06 &27.7 \pm 0.1 &602.2^{+27.7}_{-25.4} &548.8^{+28.4}_{-26.0} \pm 46.1 &11.9^{+0.3}_{-0.3} \pm 0.5&22.7\\
3.808 &50.5 &1.06 &29.1 \pm0.2 &6.2^{+4.0}_{-1.7}~($<$13.2) &312.2^{+225.6}_{-95.9} \pm35.9~($<$664.7) &9.1^{+3.3}_{-1.4} \pm0.5~($<$13.3)&2.2\\
3.867 &108.9 &1.05 &24.9 \pm 0.1 &11.0^{+3.8}_{-3.1} &302.1^{+116.9}_{-95.4} \pm 22.7 &9.1^{+1.8}_{-1.4} \pm 0.3&3.2\\
3.871 &110.3 &1.05 &24.9 \pm0.1 &7.0^{+3.3}_{-2.7}~($<$13.4) &190.3^{+100.5}_{-82.2} \pm14.3~($<$364.4) &7.2^{+1.9}_{-1.6} \pm0.3~($<$10.0)&2.3\\
3.896 &52.6 &1.05 &25.1 \pm0.1 &1.8^{+2.0}_{-0.9}~($<$5.4) &101.9^{+126.7}_{-57.0} \pm7.6~($<$305.6) &5.3^{+3.3}_{-1.5} \pm0.2~($<$9.3)&1.2\\
4.128 &401.5 &1.05 &22.6 \pm 0.1 &15.2^{+5.4}_{-3.1} &124.9^{+49.7}_{-28.5} \pm 9.4 &6.4^{+1.3}_{-0.7} \pm 0.2&3.6\\
4.157 &408.7 &1.05 &22.6 \pm0.1 &10.8^{+4.3}_{-3.2}~($<$18.8) &86.9^{+38.7}_{-28.8} \pm6.5~($<$151.2) &5.4^{+1.2}_{-0.9} \pm0.2~($<$7.1)&2.9\\
4.288 &502.4 &1.05 &21.2 \pm 0.1 &16.8^{+4.9}_{-3.8} &117.3^{+38.3}_{-29.7} \pm 8.8 &6.5^{+1.1}_{-0.8} \pm 0.2&3.8\\
4.312 &501.2 &1.05 &21.3 \pm 0.1 &15.5^{+5.0}_{-3.3} &107.9^{+39.0}_{-25.8} \pm 8.1 &6.3^{+1.1}_{-0.8} \pm 0.2&3.8\\
4.337 &505.0 &1.05 &21.3 \pm 0.1 &14.2^{+5.1}_{-2.9} &98.6^{+39.7}_{-22.6} \pm 7.4 &6.1^{+1.2}_{-0.7} \pm 0.2&3.6\\
4.377 &522.7 &1.05 &20.9 \pm0.1 &9.5^{+4.2}_{-2.5}~($<$16.8) &64.8^{+32.1}_{-19.1} \pm4.9~($<$114.5) &5.0^{+1.2}_{-0.7} \pm0.2~($<$6.6)&2.9\\
4.396 &507.8 &1.05 &20.5 \pm 0.1 &10.5^{+4.2}_{-2.5} &75.2^{+33.7}_{-20.1} \pm 5.6 &5.4^{+1.2}_{-0.7} \pm 0.2&3.2\\
4.436 &569.9 &1.05 &20.5 \pm0.2 &8.0^{+3.3}_{-2.7}~($<$14.3) &51.0^{+23.6}_{-19.3} \pm3.8~($<$91.2) &4.5^{+1.0}_{-0.8} \pm0.2~($<$6.0)&2.7\\
4.612 &103.8 &1.05 &17.7 \pm0.2 &0.5^{+1.9}_{-0.2}~($<$3.7) &20.2^{+86.1}_{-9.1} \pm1.5~($<$149.6) &3.0^{+6.4}_{-0.7} \pm0.1~($<$8.1)&0.5\\
4.628 &521.5 &1.05 &17.4 \pm0.2 &2.8^{+2.5}_{-1.5}~($<$7.5) &22.9^{+22.9}_{-13.8} \pm1.7~($<$61.5) &3.2^{+1.6}_{-1.0} \pm0.1~($<$5.2)&1.4\\
4.641 &552.4 &1.05 &17.3 \pm0.1 &6.0^{+2.8}_{-2.1}~($<$11.1) &46.5^{+24.3}_{-18.2} \pm3.5~($<$86.1) &4.6^{+1.2}_{-0.9} \pm0.2~($<$6.2)&2.4\\
4.661 &529.6 &1.05 &17.0 \pm0.1 &8.0^{+3.3}_{-2.7}~($<$14.3) &66.2^{+30.6}_{-25.0} \pm5.0~($<$118.3) &5.5^{+1.3}_{-1.0} \pm0.2~($<$7.3)&2.7\\
4.682 &1669.3 &1.05 &16.7 \pm 0.1 &12.2^{+5.1}_{-2.9} &32.5^{+15.2}_{-8.7} \pm 2.4 &3.9^{+0.9}_{-0.5} \pm 0.1&3.1\\
4.699 &536.5 &1.05 &16.8 \pm0.1 &5.5^{+3.3}_{-1.6}~($<$11.1) &45.2^{+30.4}_{-14.7} \pm3.4~($<$91.3) &4.6^{+1.5}_{-0.7} \pm0.2~($<$6.5)&2.2\\
4.781 &512.8 &1.06 &17.3 \pm0.2 &0.5^{+2.3}_{-0.5}~($<$4.6) &4.2^{+21.5}_{-4.7} \pm0.3~($<$38.5) &1.4^{+3.7}_{-0.8} \pm0.1~($<$4.3)&0.4\\
4.843 &527.3 &1.06 &16.5 \pm0.2 &-0.8^{+1.7}_{-0.8}~($<$2.0) &-6.8^{+16.2}_{-7.6} \pm0.5~($<$17.0) &0.0^{+0.0}_{-0.0} \pm0.0~($<$2.9)&0.0\\

\hline\hline
\end{tabular}}
\label{tab:signal:yields:DD}
\end{center}
\end{table}

\section{Systematic uncertainty}
The systematic uncertainties on the Born cross section measurement mainly originate from integrated luminosity, $\Xi^{-}$ reconstruction, background, angular distribution, branching fraction, and input line shape. All of these sources of systematic uncertainty are discussed in detail below.

\subsection{Luminosity}
The integrated luminosity is measured using 
$e^+e^-\ar(\gamma)e^+e^-$ events with a method similar to Refs.~\cite{BESIII:2015qfd,BESIII:2022dxl,BESIII:2022ulv} with an uncertainty of 1.0\%.

\subsection{$\Xi^{-}$ reconstruction}
The systematic uncertainty due to the $\Xi^{-}$ reconstruction efficiency incorporating the tracking and PID efficiencies, the $\Lambda$ reconstruction efficiency, the selection on $\Lambda$ and $\Xi^{-}$ decay lengths, and the $\Lambda$ and $\Xi^{-}$ mass windows, is estimated from a control sample of $\psi(3686)\rightarrow\XXB$ decays with the same method as described in Refs.~\cite{BESIII:2016nix,BESIII:2016ssr, BESIII:2019dve, BESIII:2020ktn, BESIII:2021aer, BESIII:2021gca, BESIII:2022mfx, BESIII:2022lsz, BESIII:2023lkg}. The $\Xi^{-}$ reconstruction efficiency is defined as the ratio of the number of events from the double baryon-tag $\XXB$ to that from the single baryon-tag. The difference in the $\Xi^{-}$ reconstruction efficiency between data and MC simulation is taken as the systematic uncertainty.

\subsection{Background} 
The systematic uncertainty associated with the background, which is estimated based on the sideband strategy, is determined by changing the gap between the signal region and the sideband regions from half box size to the full box size.
Since the statistics for each energy points are very limited. To avoid the statistical effect, all energy points are combined when we estimate this systematic uncertainty.
The difference of signal yields before and after changing the gap of the sideband box is taken as the systematic uncertainty.

\subsection{Angular distribution} 
In this analysis, not enough statistics is available to determine the angular distribution parameter for each energy point separately. Thus, the selection efficiency for $\EE\ar\XXB$ is determined based on a PHSP model,
which may differ from the real angular distribution. Alternatively, we utilize the joint angular distribution obtained from Ref.~\cite{BESIII:2022lsz} to reproduce a MC sample, and take the efficiency difference between the signal MC samples and the alternative MC as the systematic uncertainty due to the angular distribution.

\subsection{Branching fraction} 
The uncertainty from the branching fraction of $\Lambda\ar p\pi^{-}$ is 0.8\%, taken from the world average~\cite{PDG2020}. The uncertainty of the branching fraction of $\Xi^{-}\ar\pi^{-}\Lambda$ is negligible.

\subsection{Input line shape}
The ISR correction and the detection efficiency depend on the input line shape of the cross section. The associated systematic uncertainty  can be divided into two parts.
One part is due to the statistical uncertainty of the input line shape of the cross section, which is estimated by varying the central value of the nominal input line shape within $\pm1\sigma$ of statistical uncertainty. Then, the $(1 + \delta)\epsilon$ values for each energy point are recalculated. This process is repeated 3000 times, after which a Gaussian function is used to fit the $(1 + \delta)\epsilon$ distribution. The width of the Gaussian function is taken as the corresponding systematic uncertainty. 
The other uncertainty arises due to the resonance parameters which are fixed in the fit to the input cross section.
The resonance parameters are changed in the fit by $\pm 1\sigma$ and
the resulting changes of the values of  $(1 + \delta)\epsilon$ are taken as the systematic uncertainty. 
The total systematic uncertainty is calculated by adding the two uncertainties in quadrature. 

\subsection{Total systematic uncertainty}
The various systematic uncertainties on the Born cross section measurement are summarized in Table~\ref{systematic}.
Assuming all sources are independent, the total systematic uncertainty on the cross section measurement is determined by adding these sources in quadrature.

\begin{table}[!hpt]
	\begin{center}
	{\caption{Systematic uncertainties (in \%) and their sources for each energy point on the Born cross section measurement. Here, AD denotes angular distribution, ${\cal{B}}$ denotes branching fraction, and ILS denotes input line shape. The systematic uncertainty, which is the same for each energy point, is simplified into a single numerical value.}
        \label{systematic}
	}
	   \resizebox{1.0\columnwidth}{!}{
   \begin{tabular}{ccccccccc} \hline\hline
	$\sqrt{s}$ (GeV) &Luminosity &$\Xi^{-}$ reconstruction &Background  &AD &${\cal{B}}$  &ILS &Total     \\ \hline

3.510 & & & & & &1.2 &7.6 \\
3.650 & & & & & &1.1 &7.5 \\
3.773 & & & & & &3.8 &8.4 \\
3.808 & & & & & &8.7 &11.5 \\
3.867 & & & & & &0.9 &7.5 \\
3.871 & & & & & &1.1 &7.5 \\
3.896 & & & & & &0.8 &7.5 \\
4.128 & & & & & &0.6 &7.5 \\
4.157 & & & & & &0.4 &7.5 \\
4.288 & & & & & &1.1 &7.5 \\
4.312 & & & & & &0.6 &7.5 \\
4.337 &1.0 &5.7 &1.7 &4.3 &0.8 &0.2 &7.5 \\
4.377 & & & & & &0.5 &7.5 \\
4.396 & & & & & &0.5 &7.5 \\
4.436 & & & & & &1.0 &7.5 \\
4.612 & & & & & &1.3 &7.6 \\
4.628 & & & & & &1.0 &7.5 \\
4.641 & & & & & &0.3 &7.5 \\
4.661 & & & & & &0.4 &7.5 \\
4.682 & & & & & &0.1 &7.5 \\
4.699 & & & & & &1.5 &7.6 \\
4.781 & & & & & &0.6 &7.5 \\
4.843 & & & & & &2.2 &7.8 \\

			\hline\hline
			\end{tabular}
				}
	\end{center}
\end{table}

\section{Fit to the dressed cross section}
Potential resonances in the line shape of the cross section for the reaction $\EE\ar\XXB$ are studied by applying a fit to the dressed cross section, $\sigma^{\rm dressed} =\sigma^{B}/|1-\Pi|^2$ (without the VP effect) with the least square method. The fit minimizes
  \begin{equation}
\chi^{2} = \Delta X^{T}V^{-1}\Delta X,
\end{equation}
where $\Delta X$ is the vector of residuals between measured and fitted cross section. The covariance matrix $V$ incorporates
the correlated and uncorrelated uncertainties among different energy points, where the systematic uncertainties due to the luminosity, $\Xi^{-}$ reconstruction and branching fraction are assumed to be fully correlated among CM energies and the other ones are uncorrelated.

\begin{figure}[!hbpt]
	\begin{center}
        \includegraphics[width=0.48\textwidth]{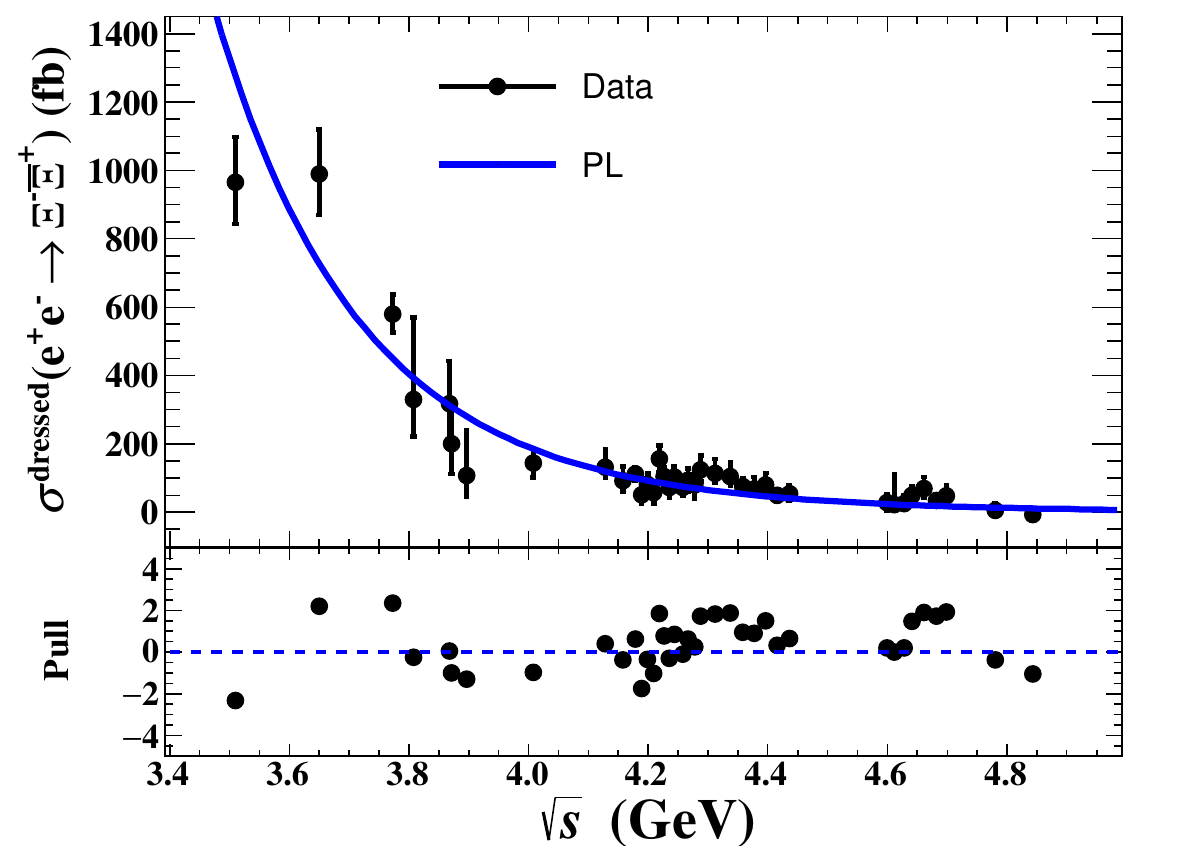}
	\includegraphics[width=0.48\textwidth]{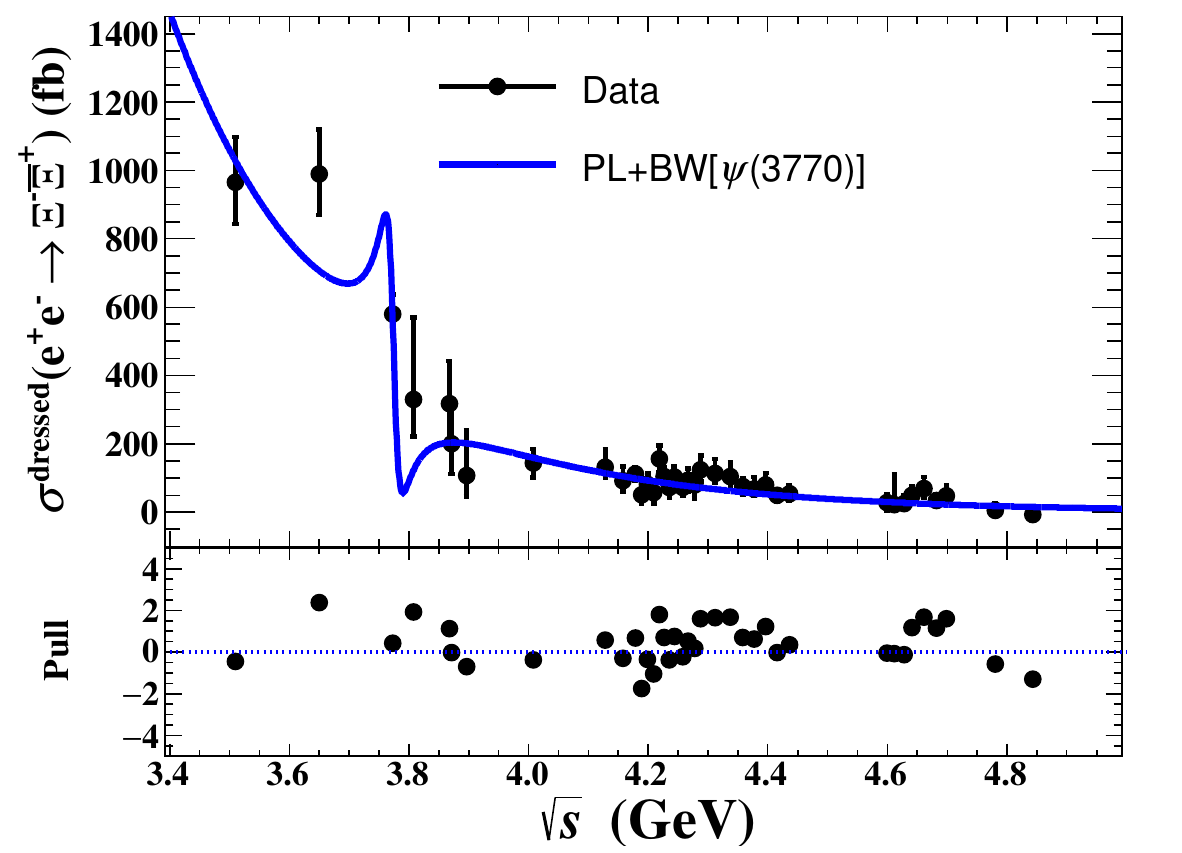}\\
 	\includegraphics[width=0.48\textwidth]{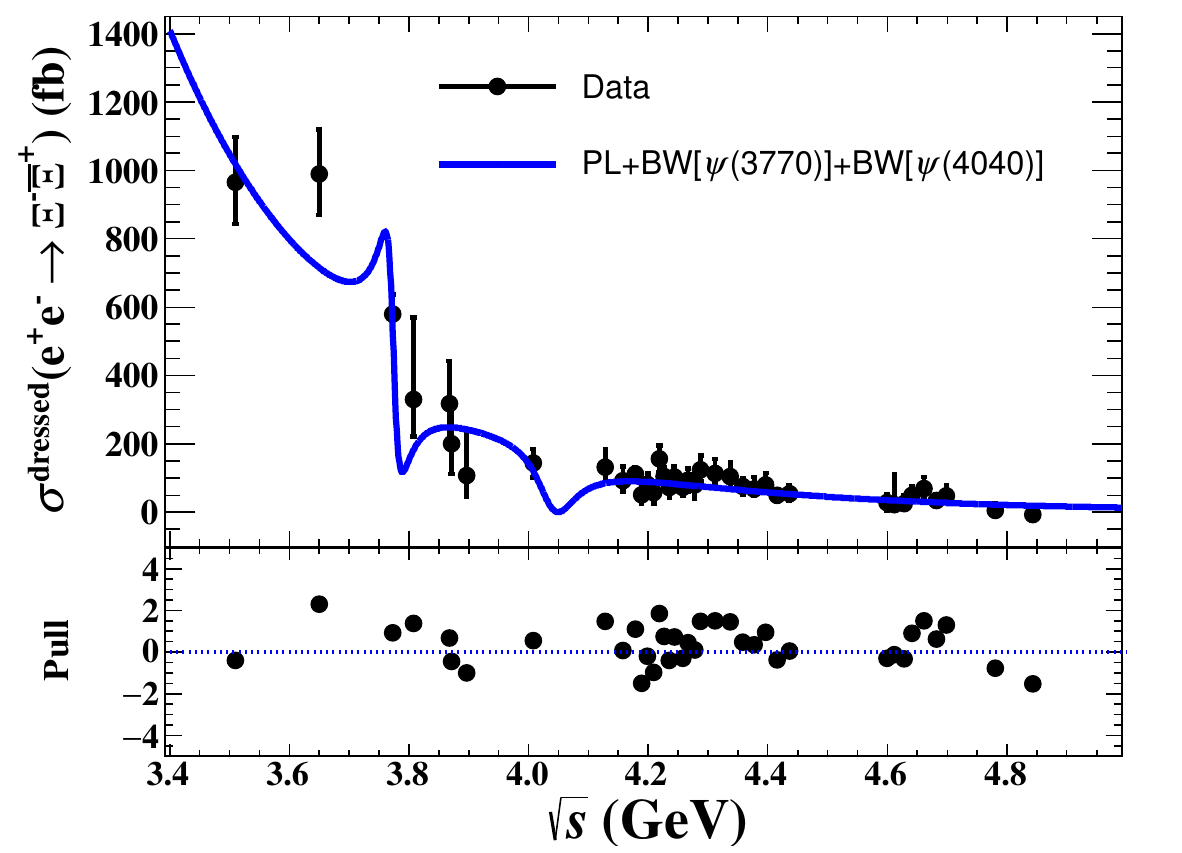}
	\includegraphics[width=0.48\textwidth]{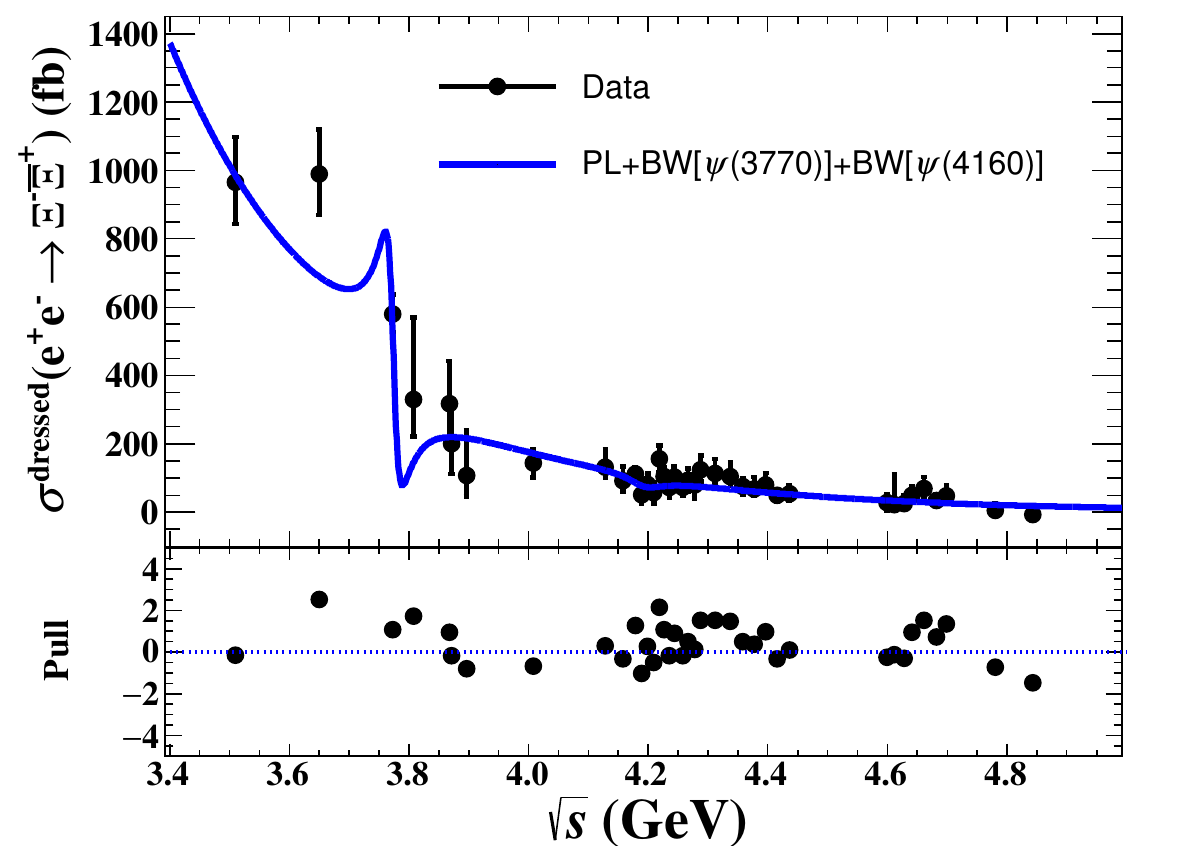}\\
 	\includegraphics[width=0.48\textwidth]{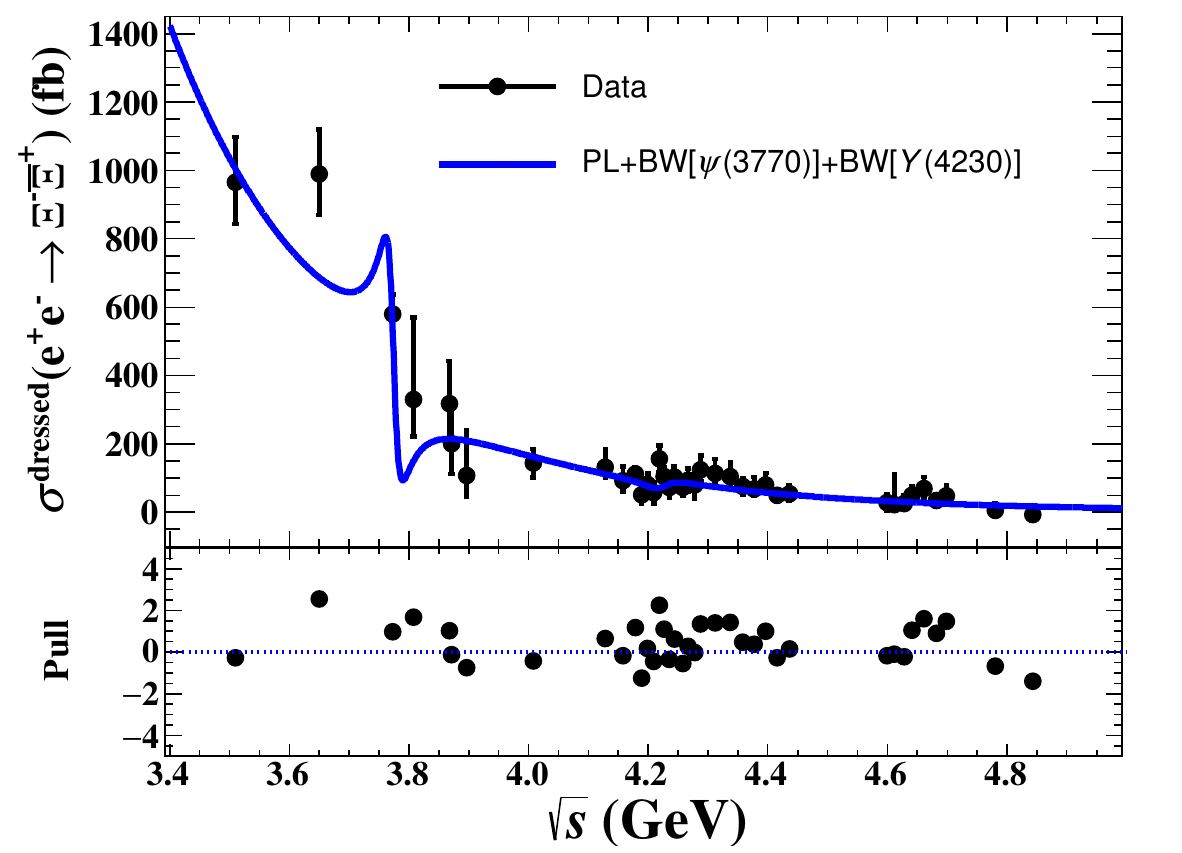}
	\includegraphics[width=0.48\textwidth]{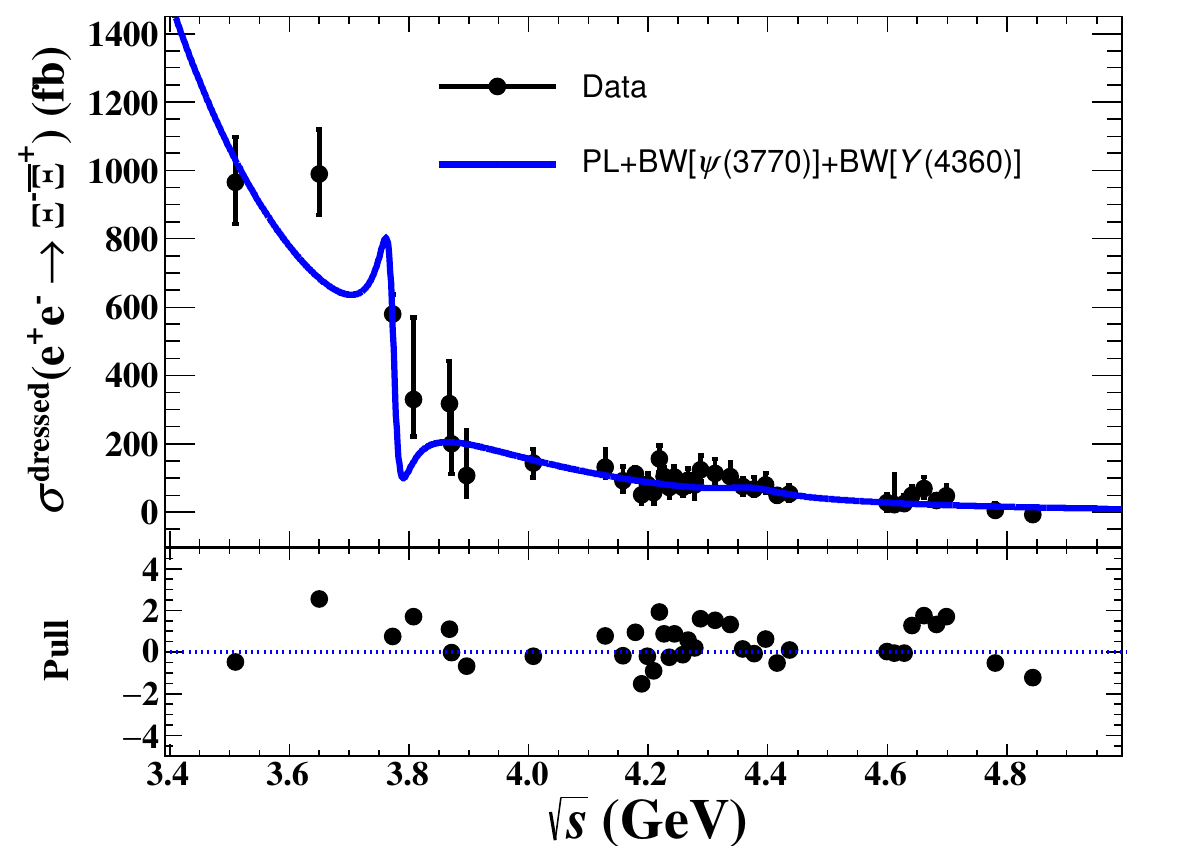}\\
 	\includegraphics[width=0.48\textwidth]{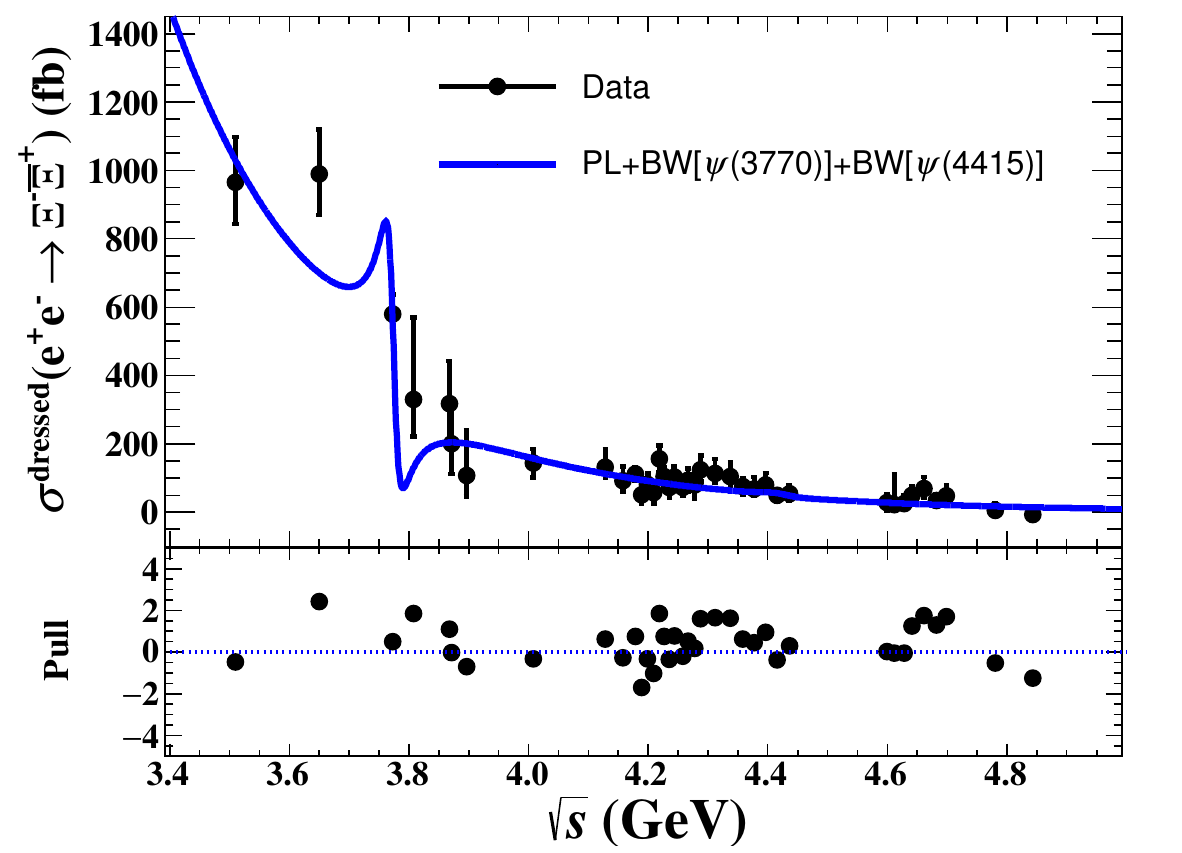}
	\includegraphics[width=0.48\textwidth]{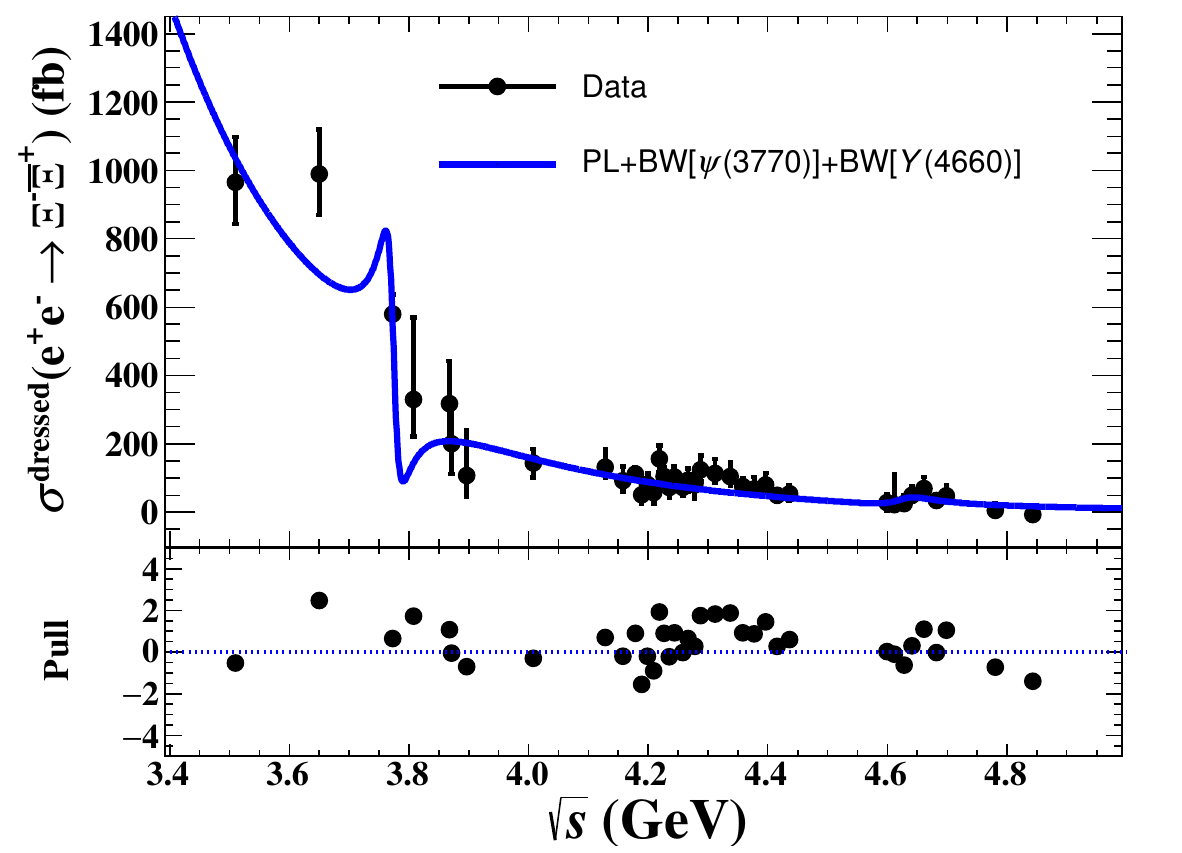}
	\end{center}
	\caption{Fits to the dressed cross section at the CM energy from 3.510 to \SI{4.843}{GeV} with the assumptions of a PL function 
only (first one), a PL function plus a ($\psi(3770)$) resonance, a PL function plus a ($\psi(3770)$) resonance and one more additional 
resonance ($\psi(4040)$, $\psi(4160)$, $Y(4230)$, $Y(4360)$, $\psi(4415)$, and $Y(4660)$). Dots with error bars are the dressed cross sections and the solid lines show the fit results.
}
	\label{Fig:XiXi::CS::Line-shape-3773}
\end{figure}

Assuming the cross section of $\EE\ar\XXB$ includes a resonance plus a continuum contribution, a fit to the dressed cross section with the coherent sum of a power-law (PL) function and a Breit-Wigner (BW) function
 \begin{equation}\label{BCS_1}
	\sigma^{\rm dressed}(\sqrt{s})= \left|c_{0}\frac{\sqrt{P(\sqrt{s})}}{\sqrt{s}^{n}} + e^{i\phi}{\rm BW}(\sqrt{s})\sqrt{\frac{P(\sqrt{s})}{P(M)}}\right|^{2},
 \end{equation}
is applied. Here $\phi$ is the relative phase between the BW function
  \begin{equation}
{\rm BW}(\sqrt{s}) =\frac{\sqrt{12\pi\Gamma_{ee}{\cal{B}}\Gamma}}{s-M^{2}+iM\Gamma},
 \end{equation}
 and the PL function, $c_0$ and $n$ are free fit parameters, $\sqrt{P(\sqrt{s})}$ is the two-body phase space factor, the mass $M$ and total width $\Gamma$ are fixed to the assumed resonance with the PDG values~\cite{PDG2020}, and $\Gamma_{ee}{\cal{B}}$ is the product of the electronic partial width and the branching fraction for the assumed resonance decaying into the $\XXB$ final state.
A fit without resonance contribution results in parameters $(c_{0} = 1.8\pm 0.4, n = 8.2 \pm 0.2)$ with the goodness of fit 
$\chi^{2}/n.d.f$ = $75.5/(38-2)$. 
The result of the fit with $\psi(3770)$ plus a continuum contribution is presented in Table~\ref{tab:multisolution}. 
Due to ambiguities in the fit, multiple solutions with the same likelihood are found~\cite{Bai:2019jrb}. Both solutions are presented in Table~\ref{tab:multisolution} under the inclusion of different resonant states.
The result of the fit with the $\psi(3770)$ plus a continuum contribution is presented in~Table~\ref{tab:multisolution}. Since the two solutions of this fit are very close, only a single solution is considered. When the other possible charmonium(- like) states, $\psi(4040)$, $\psi(4160)$, $Y(4230)$, $Y(4360)$, $\psi(4415)$, $Y(4660)$ are included one-by-one in the fit, in addition to the $\psi(3770)$  plus continuum, the result of the $\psi(3770)$ is nearly unchanged. Therefore, only the results corresponding to the additional charmonium(-like) states are presented in Table~\ref{tab:multisolution}.
Taking systematic uncertainties into account, the significance of the resonance contribution is calculated by comparing 
$\chi^{2}/n.d.f$ with and without the resonance assumption. For different assumptions, $n$ is almost unchanged. 
Evidence for the decay $\psi(3770)\ar\XXB$ with a significance of 4.5$\sigma$ including the systematic uncertainty is found 
by combining these results with earlier BESIII measurements~\cite{Ablikim:2019kkp}. 
For other possible charmonium(-like) states, no obvious signal is found and the upper limits on the products of branching fraction and two-electronic partial width 
for these charmonium(-like) states decaying into the $\XXB$ final state including the systematic uncertainty are provided 
at the 90\% confidence level (C.L.) using a Bayesian approach~\cite{Zhu:2008ca}.
Systematic uncertainties due to beam energy, mass, and width of the $\psi(3770)$ resonance have been considered
by varying the known values within one standard deviation; they turn out to be negligible. 
Figure~\ref{Fig:XiXi::CS::Line-shape-3773}
shows the fit to the dressed cross section with and without assumption of resonance contributions.

\begin{table*}[htbp]
	\centering
	\caption{The fitted resonance parameters to the dressed cross section for the $\EE\to\XXB$ process for the two ambiguous solutions.
The fit procedure includes both statistical and systematic uncertainties except for the CM energy calibration. The top row represents the result for $\psi(3770)$ plus a continuum. Only one solution is found in this case. The other rows represent the results of the resonances $\psi(4040)$, $\psi(4160)$, $Y(4230)$, $Y(4360)$, $\psi(4415)$, and $Y(4660)$, respectively, when they are added one-by-one to the fit in addition to the $\psi(3770)$ plus a continuum. 
The ${\cal{B}}$ is the branching fraction for the assumed resonance decaying into $\XXB$ final state, where possible, ${\cal{B}}$ is determined using $\Gamma_{ee}$ values from PDG \cite{PDG2020}. 
}  
 \begin{adjustbox}{scale=0.9}
        \begin{tabular}{l r@{ }l r@{ }l r@{ }l}\hline
		\multicolumn{1}{c}{Resonance parameter} &\multicolumn{2}{c}{Solution I} &\multicolumn{2}{c}{Solution II}  &\multicolumn{2}{l}{$\chi^{2}/n.d.f$}\\		\hline

    $\phi_{\psi(3770)}$ (rad)                               &$-2.1$&$\pm$ 0.2 &$-$& \\ 
    $\Gamma_{ee}\mathcal{B}_{\psi(3770)}$ ($10^{-3}$ eV)  &$35.5$&$\pm$ 9.2 &$-$& &$45.0/(38-4)$&\\ 
    ${\mathcal{B}}[\psi(3770)\to\XXB]$ ($10^{-6}$)                      &$136.0$&$\pm$ 35.2 &$-$&\\ \hline

$\phi_{\psi(4040)}$ (rad) &$-1.9 $&$\pm$ 0.2 &$-2.5 $&$\pm$ 0.1 \\
$\Gamma_{ee}\mathcal{B}_{\psi(4040)}$ ($10^{-3}$ eV) &$15.2  $&$\pm$ 27.6 $(<44.0)$ &$19.7 $&$\pm$ 30.9  $(<51.9 )$ &$37.1/(38-6)$&\\
${\mathcal{B}}[\psi(4040)\to\XXB]$ ($10^{-6}$) &$17.8$&$\pm$ 32.2 $(<51.4)$ &$23.0$&$\pm$ 36.1 $(<60.6)$ \\ \hline

$\phi_{\psi(4160)}$ (rad) &$-1.7 $&$\pm$ 0.1 &$-2.3 $&$\pm$ 0.1 \\
$\Gamma_{ee}\mathcal{B}_{\psi(4160)}$ ($10^{-3}$ eV) &$29.8  $&$\pm$ 2.5 $(<32.9)$ &$33.9 $&$\pm$ 2.7  $(<37.2 )$ &$38.1/(38-6)$&\\
${\mathcal{B}}[\psi(4160)\to\XXB]$ ($10^{-6}$) &$61.7$&$\pm$ 5.2 $(<68.1)$ &$70.2$&$\pm$ 5.6 $(<77.0)$ \\ \hline

$\phi_{Y(4230)}$ (rad) &$-1.7 $&$\pm$ 0.1 &$-2.2 $&$\pm$ 0.1 \\
$\Gamma_{ee}\mathcal{B}_{Y(4230)}$ ($10^{-3}$ eV) &$19.4  $&$\pm$ 1.9 $(<22.3)$ &$22.0 $&$\pm$ 2.1  $(<25.1 )$ &$39.5/(38-6)$&\\
${\mathcal{B}}[Y(4230)\to\XXB]$ ($10^{-6}$) &$-$& &$-$&\\ \hline

$\phi_{Y(4360)}$ (rad) &$-1.8 $&$\pm$ 0.1 &$-2.1 $&$\pm$ 0.1 \\
$\Gamma_{ee}\mathcal{B}_{Y(4360)}$ ($10^{-3}$ eV) &$36.0  $&$\pm$ 3.2 $(<41.2)$ &$39.4 $&$\pm$ 3.3  $(<44.8 )$ &$41.7/(38-6)$&\\
${\mathcal{B}}[Y(4360)\to\XXB]$ ($10^{-6}$) &$-$& &$-$&\\ \hline

$\phi_{\psi(4415)}$ (rad) &$-1.7 $&$\pm$ 0.1 &$-2.2 $&$\pm$ 0.1 \\
$\Gamma_{ee}\mathcal{B}_{\psi(4415)}$ ($10^{-3}$ eV) &$16.5  $&$\pm$ 1.9 $(<19.8)$ &$18.3 $&$\pm$ 2.0  $(<21.7 )$ &$44.5/(38-6)$&\\
${\mathcal{B}}[\psi(4415)\to\XXB]$ ($10^{-6}$) &$28.3$&$\pm$ 3.3 $(<34.0)$ &$31.4$&$\pm$ 3.4 $(<37.2)$ \\ \hline

$\phi_{Y(4660)}$ (rad) &$-1.6 $&$\pm$ 0.1 &$-2.2 $&$\pm$ 0.1 \\
$\Gamma_{ee}\mathcal{B}_{Y(4660)}$ ($10^{-3}$ eV) &$13.6  $&$\pm$ 2.0 $(<18.0)$ &$15.3 $&$\pm$ 2.2  $(<19.9 )$ &$41.1/(38-6)$&\\
${\mathcal{B}}[Y(4660)\to\XXB]$ ($10^{-6}$) &$-$& &$-$&\\ \hline

		\hline
	\end{tabular}
	\label{tab:multisolution}
 \end{adjustbox}
\end{table*}

\section{Summary}
Using a total of \SI{12.9}{fb^{-1}} of $\EE$ collision data above the open-charm threshold collected with the
BESIII detector at the BEPCII collider, the process $\EE\ar\XXB$ is studied based on a single baryon-tag technique. The Born cross sections and effective form factors are measured at 23 CM energies in the range from 3.510 to \SI{4.843}{GeV}.
Combined with earlier BESIII measurements~\cite{Ablikim:2019kkp}, a fit to the dressed cross section of the reaction $\EE\ar\XXB$ is performed, in which the line shape is described by a $\psi(3770)$ plus a continuum contribution. First evidence for the decay $\psi(3770)\ar\XXB$ is found with a significance of 4.5$\sigma$, accounting also for the systematic uncertainty. 
The branching fraction is determined to be ${\cal{B}}(\psi(3770)\ar\XXB) = (136.0\pm35.2)\times 10^{-6}$,
which is larger by at least an order of magnitude than the prediction ($4\times 10^{-7}$) based on a scaling from the electronic branching fraction values using Eq.(1) in Ref.~\cite{Dobbs:2017hyd}. This implies that the $\psi(3770)$ resonance needs to be considered when interpreting the CLEO-c data. No obvious signal from $\psi(4040)$, $\psi(4160)$, $Y(4230)$, $Y(4360)$, $\psi(4415)$, or $Y(4660)$ has been found. Thus, the upper limits of the products of branching fraction and electronic partial width for these charmonium(-like) states decaying into the $\XXB$ final state are provided at the 90\% C.L.
This measurement provides evidence of non-$D\bar{D}$ decays of $\psi(3770)$, and can be useful for the understanding of the charmonium(-like) states coupling to the baryon-antibaryon final states.  It gives additional insight into the puzzle of a large non-$D\bar{D}$ component of $\psi(3770)$ state. 

\acknowledgments
The BESIII Collaboration thanks the staff of BEPCII and the IHEP computing center for their strong support. This work is supported in part by National Key R\&D Program of China under Contracts Nos. 2020YFA0406400, 2020YFA0406300; National Natural Science Foundation of China (NSFC) under Contracts Nos. 12075107, 12247101, 11635010, 11735014, 11835012, 11935015, 11935016, 11935018, 11961141012, 12022510, 12025502, 12035009, 12035013, 12061131003, 12192260, 12192261, 12192262, 12192263, 12192264, 12192265, 12221005, 12225509, 12235017; the Chinese Academy of Sciences (CAS) Large-Scale Scientific Facility Program; the CAS Center for Excellence in Particle Physics (CCEPP); Joint Large-Scale Scientific Facility Funds of the NSFC and CAS under Contract No. U1832207; CAS Key Research Program of Frontier Sciences under Contracts Nos. QYZDJ-SSW-SLH003, QYZDJ-SSW-SLH040; 100 Talents Program of CAS; The Institute of Nuclear and Particle Physics (INPAC) and Shanghai Key Laboratory for Particle Physics and Cosmology; ERC under Contract No. 758462; European Union's Horizon 2020 research and innovation programme under Marie Sklodowska-Curie grant agreement under Contract No. 894790; German Research Foundation DFG under Contracts Nos. 443159800, 455635585, Collaborative Research Center CRC 1044, FOR5327, GRK 2149; Istituto Nazionale di Fisica Nucleare, Italy; Ministry of Development of Turkey under Contract No. DPT2006K-120470; National Research Foundation of Korea under Contract No. NRF-2022R1A2C1092335; National Science and Technology fund of Mongolia; National Science Research and Innovation Fund (NSRF) via the Program Management Unit for Human Resources \& Institutional Development, Research and Innovation of Thailand under Contract No. B16F640076; Polish National Science Centre under Contract No. 2019/35/O/ST2/02907; The Swedish Research Council; U. S. Department of Energy under Contract No. DE-FG02-05ER41374; The Olle Engkvist Foundation under contract No. 200-0605.

\end{document}